# Topological flat bands in frustrated kagome lattice CoSn

Mingu Kang,[1] Shiang Fang,[2,3] Linda Ye,[1] Hoi Chun Po,[1] Jonathan Denlinger,[4] Chris Jozwiak,[4] Aaron Bostwick,[4] Eli Rotenberg,[4] Efthimios Kaxiras,[2,3] Joseph G. Checkelsky,[1] & Riccardo Comin[1†]

[1]Department of Physics, Massachusetts Institute of Technology, Cambridge, Massachusetts 02139, USA.

[2]Department of Physics, Harvard University, Cambridge, Massachusetts 02138, USA.

[3]John A. Paulson School of Engineering and Applied Science, Harvard University, Cambridge, Massachusetts 02138, USA.

[4]Advanced Light Source, E. O. Lawrence Berkeley National Laboratory, Berkeley, California 94720, USA.

[†]rcomin@mit.edu

**Electronic flat bands in momentum space, arising from strong localization of electrons in real space, are an ideal stage to realize strong correlation phenomena. In certain lattices with built-in geometrical frustration, electronic confinement and flat bands can naturally arise from the destructive interference of electronic hopping pathways. Such lattice-borne flat bands are often endowed with nontrivial topology if combined with spin-orbit coupling, while their experimental realization in condensed matter system has been elusive so far. Here, we report the direct observation of topological flat bands in the vicinity of the Fermi level in frustrated kagome system CoSn, using angle-resolved photoemission spectroscopy and band structure calculations. The flat band manifests itself as a dispersionless electronic excitation along the G-M high symmetry direction, with an order of magnitude lower bandwidth (below 150 meV) compared to the Dirac bands originating from the same orbitals. The frustration-driven nature of the flat band is directly confirmed by the real-space chiral *d*-orbital texture of the corresponding effective Wannier wave functions. Spin-orbit coupling opens a large gap of 80 meV at the quadratic band touching point between the Dirac and flat bands, endowing a nonzero $Z_2$ topological invariant to the flat band in the two-dimensional Brillouin zone. Our observation of lattice-driven topological flat band opens a promising route to engineer novel emergent phases of matter at the crossroad between strong correlation physics and electronic topology.**

Electronic correlations are a hallmark of condensed matter systems with many-body character. Localizing electrons in real space is often considered as a route to enhance correlation effects and engineer emergent phases of matter. Most well-known examples include *d*-electron systems, where the subtle balance between kinetic energy and localization-enhanced Coulomb interaction leads to collective electron behavior and rich many-body physics encompassing unconventional superconductivity, metal-insulator transitions, density-wave instabilities, and quantum spin liquids.[1] In band-like systems, electrons can still be confined in real space in lattices supporting dispersionless electronic excitations (i.e. flat bands) in momentum space. Due to the prominence of the interaction energy scale over the quenched kinetic energy, flat bands represent a versatile platform to explore exotic correlated electron phenomena. Notable examples include *f*-electron systems with Kondo physics and heavy fermions,[2] Landau levels under high magnetic fields and the fractional quantum Hall effect,[3] and, more recently, magic-angle twisted bilayer graphene superlattices with Mott insulating phase and unconventional superconductivity.[4,5]

A known experimental route to engineering electronic confinement and flat bands relies on the destructive quantum phase interference of fermion hopping paths in certain networks including the dice, Lieb, kagome, and decorated square lattices[6-12]. Here we focus on the case of the kagome lattice depicted in Fig. 1a. In the simplest nearest neighbor electronic hopping model of the *s*-orbital kagome lattice $H = \sum_{\langle i,j \rangle} c_i^\dagger c_j + h.c.$, one can construct real space eigenfunctions with alternating phases at neighboring corners of the hexagon (Fig. 1a). This electronic state is geometrically trapped within the single hexagon since any hopping to neighboring cells is hindered by the destructive phase interference as shown in Fig. 1a. This real-space electronic localization translates into momentum-space (Bloch) eigenfunctions with no energy dispersion, namely flat bands (Fig. 1b). In the tight-binding model of kagome lattice, this dispersionless excitation materializes alongside a pair of Dirac bands that are protected by the lattice symmetry similar to the case of the honeycomb lattice. Both the linear band crossing at K and quadratic band touching point at $\Gamma$ become gapped once spin-orbit coupling (SOC) is included, and the Dirac and flat bands become topologically nontrivial.[10-15] This peculiar band structure of the kagome lattice has recently attracted significant interest, not only in the context of electronic topology – topological insulator, Chern insulator, and fractional quantum Hall phases[10-15] – but also as a platform to realize many-

body electronic phases – density waves, magnetism, Pomeranchuk instability, and superconductivity.[16-18]

Unlike the ideal case, in real materials the dispersion of the flat band is modified by additional factors such as in-plane next-nearest-neighbor hopping, interlayer coupling, and multiple orbital degrees of freedom. Thus, the identification of a prototypical kagome compound hosting pristine flat bands and the characterization of the robustness and flatness of these electronic states are key steps toward realizing strongly correlated topological phenomena in this family of materials. While these phenomena have been recently observed in optical and engineered atomic Lieb lattices, the experimental realization of the lattice-borne flat bands in a solid-state system has remained elusive.[19,20] Prior scanning tunneling microscopy (STM) studies on kagome compounds $Fe_3Sn_2$ and $Co_3Sn_2S_2$ reported the observation of flat bands in the momentum-integrated density of states.[21-23] However, detailed band structure calculations revealed that the region of existence of these dispersionless excitations is rather limited in the momentum-space in these compounds,[21,24] presumably due to complex hopping pathways in realistic kagome lattices. These calculations are consistent with the relatively weak enhancement of the density of states in STM measurements compared to the ideal flat band case.[21-23] Accordingly, ARPES experiments have been carried out on $Fe_3Sn_2$ and $Co_3Sn_2S_2$, but no *bona fide* flat bands have been detected.[24,25] The unambiguous momentum-space identification of the kagome-based flat band and analysis of its topological character have, therefore, remained the subject of ongoing investigations.

In the present study of kagome metal CoSn, we combine ARPES and band structure calculations to report the presence of topological flat bands with suppressed dispersion in all three momentum space directions. CoSn belongs to the family of binary kagome metals $T_mX_n$ ($T$ : 3$d$ transition metals, $X$ : Sn, Ge, $m$:$n$=3:1, 3:2, 1:1), wherein the kagome network is constructed upon 3$d$ transition metals as shown in Fig. 1c. Compounded with the variety of magnetic ground states and topological electronic structures, this materials family has been recently spotlighted as a versatile platform for novel correlated topological phases.[25-29] For example, previous studies on $Mn_3$(Sn/Ge), $Fe_3Sn_2$, and FeSn have revealed band singularities ranging from three-dimensional Weyl points to two-dimensional (2D) Dirac points, which, in combination with intrinsic magnetism, generate large and intrinsic anomalous Hall conductivity.[25-29] In FeSn, with spatially decoupled kagome planes, the kagome-derived flat band associated with the observed Dirac band structure is theoretically predicted at 0.5 eV above the Fermi level.[26] The replacement of Co at the transition metal site suppresses the formation of local moments and magnetic ordering in this compound presumably due to a higher *d*-orbital filling,[30] while, at the same time, shifts the overall band structure below the Fermi energy, so that all kagome-derived electronic excitations (including the flat band) can be accessed by ARPES. Consequently, we could directly visualize the kagome-derived flat band as well as the large SOC gap at the quadratic band touching point between Dirac and flat bands, which endows nontrivial topology to the flat band as long predicted theoretically.

Figure 2 summarizes the experimental band structure of CoSn as measured using ARPES. The data in Fig. 2a-h were acquired using 92 eV photo excitation, which maximizes the signal from kagome band structures. Figure 2a shows the Fermi surface of CoSn and its characteristic hexagonal symmetry (the surface Brillouin zone is marked by the white-dashed lines) as expected from the underlying symmetry of the kagome lattice. In Figure 2b-f we display a series of energy-momentum dispersions measured at $k_y = 0.0$, 0.40, 0.79, and 1.19 Å$^{-1}$ (corresponding to orange, brown, green, and cyan traces in Fig. 2a respectively) across various high symmetry points. As shown in Fig. 2b,c, the energy momentum dispersion along the $\bar{\Gamma} - \bar{M}$ high symmetry direction displays a strikingly nondispersive band near the Fermi level at $-0.27 \pm 0.05$ eV, which manifests itself independently of photon polarization. The dispersion of the flat band in this specific direction is lower than the experimental broadening of the band, which is below 50 meV. As shown in Fig. 2d-f, the nondispersive nature of the flat band spans most of the Brillouin zone, and acquires a small dispersion only close to the $\bar{K}$ point. We note that the acquisition of a small but finite bandwidth near the K point is typical for realistic kagome models due to next-nearest-neighbor hopping.[14,22,31] Even using conservative estimates, the bandwidth of the flat band over the entire Brillouin zone does not exceed 150 meV, suggesting that the electron kinetic energy is strongly quenched by quantum interference effects, preventing delocalization of the wave function across the lattice.

To further examine the nondispersiveness of the flat band, we present in Fig. 2g the experimental band structure measured along the $\bar{\Gamma} - \bar{K} - \bar{M}$ high symmetry direction. Near the $\bar{K}$ point, a linearly dispersing Dirac band is found (see also Fig. 2d,e and Fig. S1 for further characterization of the Dirac bands), which is also characteristic of the kagome band structure as previously observed in Fe$_3$Sn$_2$ and FeSn.[25,26] The Dirac point is located at $-0.57 \pm 0.05$ eV, and only one branch of the Dirac cone could be observed along $\bar{\Gamma} - \bar{K}$ direction due to the matrix element effect associated to the chirality of the Dirac fermion similar to graphene.[26,32] The velocity of the Dirac band is $(1.8 \pm 0.1) \times 10^6$ m/s, which is renormalized by 23 % from the DFT value (see below) similar to magnetic 3$d$ metals Fe and Ni.[33,34] With the yellow-dashed boxes and red and blue bars of Fig. 2g, we directly compare the bandwidth of the Dirac and flat bands. The bandwidth of Dirac band extends over a range of 1.5 eV, which is typical for 3$d$-electron systems such as elemental transition metals and cuprates.[35-37] In contrast, the reduced bandwidth of the flat band (< 0.15 eV) is highly unusual, and can be regarded as a direct consequence of quantum phase interference effects in the kagome lattice as introduced in Fig. 1a.

A defining trait of the flat band is a diverging density of state (DOS), which often sets the stage for emergent electronic phases characterized by collective electronic, magnetic, and superconducting orders.[16-18] The high DOS from the flat band in CoSn can be captured (if we neglect the photoemission matrix element effect) from the momentum-integrated energy distribution curves shown in Fig. 2h. Here, colored lines are the momentum-integrated spectra from the energy-momentum sections in Fig. 2c-f, while the black line is obtained by integration over the full first Brillouin zone. All integrated energy distribution curves show a sharp and intense peak near the fixed flat band energy –0.27 ± 0.05 eV), reflecting the high DOS associated to the nondispersive flat band. In contrast, the spectral weight from all other bands (including the Dirac bands) is spread out in energy and further modulated as a function of $k_y$.

In Fig. 2i and Fig. 2j-m we present the evolution of the flat band as a function of out-of-plane momentum $k_z$ as measured using photon energy-dependent ARPES. The flat band displays negligible dispersion (< 50 meV) along Γ-A as shown in Fig. 2i, while fully retaining its in-plane flatness as shown in the representative cuts of Fig. 2j-m, measured at $k_z$ = 0, π/3, 2π/3, and π (mod 2π). The flat band can thus be considered *flat* also along the $k_z$ direction. Even though the bandwidth-canceling mechanisms are different for the in-plane and out-of-plane directions – the former is quenched by quantum phase interference, while the latter is suppressed by virtue of structural layering (see Fig. 1d) – the flat dispersion is realized along all momentum directions in CoSn, at variance with the limited momentum-range of the flat bands in previously studied kagome compounds.[21-24] Accordingly, the kinetic energy of flat band electrons in CoSn is strongly quenched, and strongly correlated many-body ground states are naturally expected for partial filling of the flat band. In other words, if we start from a simple Hubbard model $H = -t \sum_{\langle i,j \rangle} c_i^\dagger c_j + U \sum_i n_{i\uparrow} n_{i\downarrow}$, where $t$ is the hopping integral and $U$ is the on-site interaction, the large $U/t$ value essential to promote strong electronic correlations can be attained even with relative small $U$ value thanks to the quenched $t$ of the flat bands. We calculate $U \approx 5$~$6$ eV in CoSn, based on the linear response approach[38], confirming that the interaction energy scale dominates both the quenched bandwidth (< 0.2 eV) and nearest-neighbor hopping strength ($t \approx 0.015$ eV) for the flat bands (see Methods and Supplementary Informations). The realization of strongly correlated phases based on the flat band electrons has been recently demonstrated in magic-angle twisted Moiré superlattices, whose flat minibands serve as a basis for Mott-insulating, superconducting, and magnetic ground states.[4,5,39] Similar correlated states of matter have been theoretically investigated for the case of the flat band in the kagome lattice.[16,17,40-43] Here, we experimentally identify a candidate material to extend this research avenue to kagome systems.

To understand the origin of the observed flat band, we use relativistic DFT to calculate the theoretical band structure of CoSn as shown in Fig. 1e. Despite the complexity inherent to the presence of multiple $d$-orbital degrees of freedom, the calculations closely capture the experimental manifestation of the kagome band structure. For example, in the $k_z = 0$ plane, a strongly dispersing Dirac bands (with Dirac points at K) is reproduced between −0.3 and −1.8 eV with a 1.1 ~ 1.5 eV bandwidth. At the same time, two flat bands (with bandwidth quenched below 0.2 eV) appear above the Dirac bands (highlighted with orange and cyan boxes), in close correspondence to Fig. 1b. Our analysis of orbital character in Fig. S2 reveals that the two flat bands arise from different orbital degrees of freedom: the upper (lower) flat band in the orange (cyan) box is mainly formed by $d_{xy}/d_{x2-y2}$ ($d_{xz}/d_{yz}$) orbitals. These flat bands also have suppressed dispersion along $k_z$ direction (see Γ-A direction for example), consistent with experimental observations. We also found a flat band at $k_z = \pi$ (brown box in Fig. 1e), which arises from $d_{z2}$ orbital. However, due to its out-of-plane orbital character, the band is not effectively localized in the $k_z$ direction, exhibiting a bandwidth > 1 eV. We note that the steep $k_z$ dispersive bands may account for the low out-of-plane resistivity and positive out-of-plane Hall coefficient observed in our transport measurements on CoSn (Fig. S3).

For a detailed comparison between experiment and theory, we tune the photon energy to 128 eV to visualize the in-plane electronic structure at the $k_z = 0$ (mod 2π) plane and acquire high-resolution energy-momentum maps of the flat band. As shown in Fig. 3a,c, we could directly detect the two flat bands with dispersion < 0.1 eV along Γ-M and < 0.2 eV along Γ-K-M. The experimental dispersion closely follows the theoretical dispersion shown in Fig. 3b,d (which are shifted up by 140 meV to match the experimental Fermi level), confirming the assignment of these features to the two flat bands arising from different $d$-orbital degrees of freedom as discussed above. The Dirac point is again observed at K, and positioned at slightly higher binding energy 0.73 ± 0.05 eV due to small but finite dispersion of the Dirac bands along $k_z$. We note that the flat band highlighted in Fig. 2 corresponds to the lower flat band (cyan) with $d_{xz}/d_{yz}$ orbital characters. Below, we will focus on these prototypical flat bands at $k_z = 0$ to analyze their localization and topology.

At this point, an important outstanding question is how the localization mechanism in the simple $s$-orbital kagome tight-binding model (Fig. 1a,b) manifests in the realistic $d$-orbital kagome lattice of CoSn. To address this aspect, we derived a DFT-based *ab initio* tight-binding model of CoSn (Fig. S4), and use the $k_z = 0$ flat bands to construct the real-space effective Wannier functions on the 2D kagome plane (see Supplementary Informations for details). We construct the flat band Wannier function to retain the highest degree of symmetries possible (except those abandoned by the topological obstructions associated with nontrivial $Z_2$ invariant and mirror Chern number; see discussion below), which include a subset of important symmetries of the kagome lattice such as $C_6$ rotational symmetry, $xz/yz$ mirror symmetry, combined inversion/time-reversal symmetry, and combined $xy$ mirror/time-reversal symmetry. As such, the Wannier functions we derived could serve as a basis for future analyses of interaction effects within the flat bands. The real-space orbital textures of the constructed flat band Wannier functions are displayed in Fig. 3e,f, while the corresponding spin textures are displayed in Fig. S5. Several important points are apparent: (1) Due to the finite dispersion of the flat bands in CoSn, we could observe a finite charge density leaking out of the central hexagon unlike the ideal case in Fig. 1a. Nonetheless, the charge density rapidly and exponentially decays away from the central hexagon (insets in Fig. 3e,f), and 85 % of total charge are confined in the first to third sites from the center. This provides the length scale of the localization ≈ 7 Å. (2) If we focus on the states at the central hexagon, chiral orbital textures could be observed around the hexagon, in which the $d$-orbitals at the neighboring corners are aligned antiphase toward the site outside of the hexagon. This is in reminiscence of the alternating phases in the compact localized states of $s$-orbital kagome model in Fig. 1a. In the Supplementary Informations, we have demonstrated that such orbital textures combined with fine-tuned multiorbital hopping parameters in the kagome lattice geometry suppress the charge leakage outside of the hexagon and localize the real-space electronic wave functions. This finding implies that the destructive quantum phase interference from alternating phases in the *s*-wave kagome tight-binding model is transferred to the real-space chiral orbital textures in realistic *d*-orbital kagome model. This analogy unequivocally confirms the frustration-driven origin of the observed flat bands in CoSn. (3) The Wannier functions centered at the hexagon of the kagome lattice effectively forms a triangular lattice. We estimate the hopping *t* between Wannier functions at neighboring triangular lattice sites to be ≈ 15 meV reflecting suppressed kinetic energy of the flat band electrons (Table S4). In this context, the exotic electronic phases recently derived from the Hubbard model on triangular lattices including spiral magnetic orders, unconventional superconductivity, and quantum or chiral spin liquids might be relevant to the flat bands in kagome lattice.[44-46]

After demonstrating the realization of the kagome-derived flat band in CoSn, we examined the spin-orbit coupling-induced gap opening at the quadratic touching point (at Γ) between the Dirac band and the flat band. In the viewpoint of the prototypical tight-binding framework for the kagome lattice, the gap at the quadratic band touching point is responsible for rendering the flat band topologically nontrivial (see Fig. 1b and Fig. 4c), endowing nonzero Chern number / $Z_2$ invariant under the time-reversal breaking / symmetric condition.[10-15] Figure 4a highlights the band dispersion near the Γ point obtained by averaging three spectra taken at the first, second, and third Brillouin zone to minimize the influence of photoemission matrix element effect to the intensity distributions. The band dispersion closely follows the DFT calculation shown in Fig. 4b, and in particular, it exhibits a quadratic band that emerges from the Dirac band at K (see also Fig. 3c,d) and touches the flat band at Γ. Focusing on this quadratic band and the flat band with same orbital origin ($d_{xz}/d_{yz}$ orbitals, cyan-colored lines Fig. 4b), the detailed analysis of the energy distribution curves in Fig. 4d clearly reveals the spin-orbit-induced gap opening at the quadratic touching point. The spin-orbit coupling gap size could be quantified to be $\Delta_{ARPES}$ = 80 ± 20 meV ($\Delta_{DFT}$ = 57 meV), which is considerably larger than the gap observed at the linear band crossing of $Fe_3Sn_2$ ≈ 30 meV.[25] The direct observation of the spin-orbit coupling gap between the Dirac and the flat band strongly signals the nontrivial topology of the observed flat band at $k_z$ = 0. To support this, we use the DFT-derived Wannier tight-binding model to analyzed of the parity eigenvalue of the flat bands at the $k_z$ = 0 plane following Fu-Kane formula,[47] which yields topological index $Z_2$ = 1 for both flat bands confirming their topological nature (Fig. 4e). We note that the nontrivial topology of the flat bands is also reflected in the spin-texture of the constructed Wannier wave functions (Fig. S5), which inevitably breaks time-reversal symmetry due to the $Z_2$ obstruction.[48] In the time-reversal symmetry breaking setting, the presence of a gap can determine bulk properties like the anomalous Hall conductivity via the Berry curvature mechanism. When extended to the 2D limit of a single kagome sheet, the 2D topological insulator state can be realized based on the topological flat band (Fig. 4f), and helical edge states are expected inside the SOC gap. Overall, our observation of the SOC gap between the Dirac and flat bands confirms the realization of another defining properties of 2D kagome band structure in CoSn, which endows nontrivial topological character to the correlated flat band electrons.

In sum, we have successfully discovered and characterized the topological flat bands in the frustrated kagome lattice CoSn. An important future endeavor would be to bring this spectroscopically identified flat band to the Fermi level to observe lattice-borne correlated topological phases. Potential routes to this include bulk doping (for example, by $Co_{1-x}Fe_xSn$), monolayer fabrication, and application of compressive in-plane strain as suggested by our DFT calculations (Fig. S6-8). Altogether, the observation of topological flat bands in the kagome lattice opens up a new avenue to study correlation-driven emergent electronic phenomena on the background of topological nontriviality.

## Methods

**Synthesis of CoSn single crystals.** Single crystals of CoSn were synthesized using a Sn self flux method. Cobalt powder (Alfa Aesar, 99.998 %) and tin pieces (Alfa Aesar, 99.9999 %) were mixed with molar ratio 1: 9 and loaded in an alumina crucible and sealed in a quartz tube under high vacuum. The tube was heated to 950 °C and maintained for 5 hours, and was gradually cooled to 650 °C with a typical cooling rate of 2-3 °C /hour. At 650 °C the tube was removed from the furnace and centrifuge dissociation was performed to separate the crystals from flux. Crystals with a hexagonal prismatic shape were obtained with the longest dimension (6-8 mm) typically along [001].

**Angle-resolved photoemission spectroscopy (ARPES).** ARPES experiments were performed at Beamline 4 (MERLIN) and Beamline 7 (MAESTRO) of the Advanced Light Source equipped with R8000 and R4000 hemispherical electron analyzers (Scienta Omicron) respectively. For ARPES measurements, surface of CoSn was prepared in two different ways, one by *ex situ* fine polishing followed by *in situ* Ar$^+$ ion-sputtering and annealing at 1000 °C and the other by *in situ* low-temperature cleaving. The experiments were performed below ≈ 60 K and under the ultrahigh vacuum ≈ 4 × 10$^{-11}$ torr. The photon energy was scanned in the range from 30 eV to 160 eV which covers more than two full three-dimensional Brillouin zone of CoSn. Corresponding $k_z$ momentum was calculated by assuming nearly-free-electron final state with inner potential 5.5 eV. We used *s*-polarized photons unless specified, which maximize the signal from the flat bands. The convoluted energy and momentum resolutions of the beamline and the analyzer were better than 30 meV and 0.01 Å$^{-1}$ respectively. Termination-dependence or the effect of surface polarity have not been observed in our ARPES experiment.

**Electronic structure calculation from the first principles.** Density functional theory (DFT) calculations were performed with the Vienna Ab initio Simulation Package (VASP).[49,50] The pseudo-potentials are of the Projector Augmented Wave (PAW)[51] type with exchangecorrelation energy functional parameterized by Perdew, Burke and Ernzerhof (PBE)[52] within the generalized gradient approximation (GGA). The bulk DFT calculations are converged with plane-wave energy cut-off 350 eV and a reciprocal space Monkhorst-Pack grid sampling of size 15 × 15 × 11. The spin-orbit coupling (SOC) terms are included for the electronic band structure. The calculation of the effective interaction parameters $U$ in CoSn is performed based on the standard linear response approach following Ref. 38 (see Fig. S13 for details). We note that the theoretical Fermi level from the DFT needs to be shifted down by 140 meV to fit the experimental band structure, presumably due to a slight Sn off-stoichiometry in our crystals. Such shifting has been applied in the calculations presented in Fig. 1e, Fig. 2h, Fig. 3b,d, Fig. 4b, Fig. S2, and Fig. S4.

**Wannier tight-binding model of CoSn.** To interpret the first principle calculations, the post-process Wannier90 code[53] is used to convert the extended periodic Bloch wave function basis in the DFT into the localized Wannier functions basis in the real space[54]. With this Wannier transformation, the effective tight-binding Hamiltonian for a selected group of bands of the material can be constructed. The construction gives both the accurate DFT bands and physically transparent pictures of localized atomic orbitals and their mutual hybridizations. Here, we have included Co $3d$ states, and Sn $5s$ and $5p$ states in the Wannier construction for the bands around the Fermi level. The Wannier model is first constructed without SOC. The SOC interactions are modeled with atomic terms which capture the essential feature of the DFT bands with SOC. With these Wannier models, the symmetry properties and the topological index could be directly computed. In Fig S4a,b, we show the band structure comparison between the full DFT calculations and the interpolated Wannier tight-binding model with and without spin-orbit coupling modeling respectively. The extracted spin-orbit coupling strength $\lambda \vec{L} \cdot \vec{S}$ has $\lambda_{Co} \approx 70$ meV for Co atoms and $\lambda_{Sn} \approx 290$ meV for Sn atoms. Hybridization between Co and Sn orbitals may account for the larger mass of the Dirac electrons at K (see Fig. S2a and Fig. S4b) compared to related kagome compounds $Fe_3Sn_2$ and $FeSn$.[25,26]

# References

1. Maekawa, S., Tohyama, T., Barnes, S. E., Ishihara, S., Koshibae, W. & Khaliulin, G. *Physics of transition metal oxides,* Springer Nature, Switzerland (2004).

2. Si, Q. & Steglich, F. Heavy fermions and quantum phase transitions, *Science* **329**, 1161-1166 (2010).

3. Tsui, D. C., Stormer, H. L. & Gossard, A. C. Two-dimensional magnetotransport in the extreme quantum limit, *Physical Review Letters* **48**, 1559-1562 (1982).

4. Cao, Y. *et al.* Correlated insulator behaviour at half-filling in magic-angle graphene superlattices, *Nature* **556**, 80–84 (2018).

5. Cao, Y. *et al.* Unconventional superconductivity in magic-angle graphene superlattices, *Nature* **556**, 43–50 (2018).

6. Sutherland, B. Localization of electronic wave functions due to local topology, *Physical Review B* **34**, 5208–5211 (1986).

7. Lieb, E. H. Two Theorems on the Hubbard Model, *Physical Review B* **62**, 1201–1204 (1989).

8. Leykam, D., Andreanov, A. & Flach, S. Artificial flat band systems: From lattice models to experiments, *Advances in Physics: X* **3**, 677–701 (2018).

9. Wu, C., Bergman, D., Balents, L. & Das Sarma, S. Flat bands and wigner crystallization in the honeycomb optical lattice, *Physical Review Letters* **99**, 070401 (2007).

10. Sun, K., Gu, Z., Katsura, H. & Das Sarma, S. Nearly Flatbands with Nontrivial Topology, *Physical Review Letters* **106**, 236803 (2011).

11. Tang, E., Mei, J.-W. & Wen, X.-G. High-Temperature Fractional Quantum Hall States, *Physical Review Letters* **106**, 236802 (2011).

12. Neupert, T., Santos, L., Chamon, C. & Mudry, C. Fractional Quantum Hall States at Zero Magnetic Field, *Physical Review Letters* **106**, 236804 (2011).

13. Guo, H. M. & Franz, M. Topological insulator on the kagome lattice, *Physical Review B* **80**, 113102 (2009).

14. Xu, G., Lian, B. & Zhang, S.-C. Intrinsic Quantum Anomalous Hall Effect in the Kagome Lattice Cs2LiMn3F12, *Physical Review Letters* **115**, 186802 (2015).

15. Bolens, A. & Nagaosa, N. Topological states on the breathing kagome lattice, *Physical Review B* **99**, 165141 (2019).

16. Mielke, A. Exact ground states for the Hubbard model on the Kagome lattice, *Journal of Physics A: Mathematical and General* **25**, 4335–4345 (1992).

17. Wen, J., Ruegg, A., Wang, C. C. & Fiete, G. A. Interaction-driven topological insulators on the kagome and the decorated honeycomb lattices, *Physical Review B* **82**, 075125 (2010).

18. Kiesel, M. L., Platt, C. & Thomale, R. Unconventional fermi surface instabilities in the kagome hubbard model, *Physical Review Letters* **110**, 126405 (2013).

19. Taie, S., Ozawa, H., Ichinose, T., Nishio, T., Nakajima, S. & Takahashi, Y. Coherent driving and freezing of bosonic matter wave in an optical Lieb lattice, *Science Advances* **1**, e1500854 (2015).
20. Drost, R., Ojanen, T., Harju, A. & Liljeroth, P. Topological states in engineered atomic lattices, *Nature Physics* **13**, 668–671 (2017).
21. Lin, Z. *et al.* Flatbands and Emergent Ferromagnetic Ordering in Fe3Sn2 Kagome Lattices, *Physical Review Letters* **121**, 096401 (2018).
22. Yin, J.-X. *et al.* Negative flat band magnetism in a spinorbit- coupled correlated kagome magnet, *Nature Physics* **15**, 443–448 (2019).
23. Jiao, L. *et al.* Signatures for half-metallicity and nontrivial surface states in the kagome lattice Weyl semimetal $Co_3Sn_2S_2$. Phys. Rev. B **99**, 245158 (2019).
24. Liu, D. F. *et al.* Magnetic Weyl semimetal phase in a Kagome crystal, *Science* **365**, 1282-1285 (2019).
25. Ye, L. *et al.* Massive Dirac fermions in a ferromagnetic kagome metal, *Nature* **555**, 638–642 (2018).
26. Kang, M. *et al.* Dirac fermions and flat bands in ideal kagome metal FeSn https://arxiv.org/abs/1906.02167 (2019).
27. Nakatsuji, S. Kiyohara, N. & Higo, T. Large anomalous Hall effect in a non-collinear antiferromagnet at room temperature, *Nature* **527**, 212–215 (2015).
28. Kuroda, K. *et al.* Evidence for Magnetic Weyl Fermions in a Correlated Metal, *Nature Materials* **16**, 1090– 1095 (2017).
29. Nayak, A. K. *et al.* Large anomalous Hall effect driven by non-vanishing Berry curvature in non-collinear antiferromagnet Mn3Ge, *Sciences Advances* **2**, e1501870 (2016).
30. Allred, J. M., Jia, S., Bremholm, M., Chan, B. C. & Cava, R. J. Ordered CoSn-type ternary phases in Co 3Sn 3-xGe x, *Journal of Alloys and Compounds* **539**, 137–143 (2012).
31. Mazin, I. I. *et al.* Theoretical prediction of a strongly correlated Dirac metal, *Nature Communications* **5**, 4261 (2014).
32. Liu, Y., Bian, G., Miller, T. & Chiang, T. C. Visualizing electronic chirality and Berry phases in graphene systems using photoemission with circularly polarized light, *Physical Review Letters* **107**, 166803 (2011).
33. Schafer, J., Hoinkis, M., Rotenberg, E., Blaha, P. & Claessen, R. Fermi surface and electron correlation effects of ferromagnetic iron, *Physical Review B* **72**, 155115 (2005).
34. Eberhardt, W. & Plummer, E. W. Angle-resolved photoemission determination of the band structure and multielectron excitations in Ni, *Physical Review B* **21**, 3245–3255 (1980).
35. Courths, R., Cord, B., Wern, H. & Hufner, S. Angle-Resolved Photoemission and Band Structure of Copper, *Physica Scripta* **1983**, 144–147 (1983).
36. Kamakura, N. *et al.* Bulk band structure and Fermi surface of nickel: A soft x-ray angle-resolved photoemission study, *Physical Review B* **74**, 045127 (2006).

37. Damascelli, A., Hussain, Z. & Shen, Z. Angle-resolved photoemission studies of the cuprate superconductors, *Reviews of Modern Physics* **75**, 473–541 (2003).

38. Cococcioni, M. & de Gironcoli, S. Linear resonse approach to the calculation of the effective interaction parameters in the LDA + U method, *Phys. Rev. B* **71**, 035105 (2005).

39. Sharpe, A. L. *et al.* Emergent ferromagnetism near three-quarters filling in twisted bilayer graphene, *Science* **365**, 605-608 (2019).

40. Chen, Y. *et al.* Ferromagnetism and Wigner crystallization in kagome graphene and related structures, *Phys. Rev. B* **98**, 035135 (2018).

41. Iglovikov, V. I. *et al.* Superconducting transitions in flat-band systems, *Phys. Rev. B* **90**, 094506 (2014).

42. Huhtinen, K. *et al.* Spin-imbalanced pairing and Fermi surface deformation in flat bands, *Phys. Rev. B* **97**, 214503 (2018).

43. Han, W. H. *et al.* A metal-insulator transition via Wigner crystallization in Boron triangular kagome lattice, https://arxiv.org/abs/1902.08390 (2019)

44. Sahebsara, P. & Senechal, D. Hubbard model on the triangular lattice: Spiral order and spin liquid, *Physical Review Letters* **100**, 136402 (2008).

45. Szasz, A., Motruk, J., Zaletel, M. P. & Moore, J. E. Observation of a chiral spin liquid phase of the Hubbard model on the triangular lattice: a density matrix renormalization group study, https://arxiv.org/abs/1808.00463 (2018).

46. Venderley, J. *et al.* A DMRG study of superconductivity in the triangular lattice Hubbard model, https://arxiv.org/abs/1901.11034 (2019).

47. Fu, L. & Kane, C. L. Topological insulators with inversion symmetry, *Physical Review B* **76**, 045302 (2007).

48. Soluyanov, A. A. & Vanderbilt, D. Wannier representation of Z2 topological insulators, *Physical Review B* **83**, 035108 (2011).

49. Kresse, G. & Furthmuller, J. Efficient iterative schemes for ab initio total-energy calculations using a plane-wave basis set, *Physical Review B* **54**, 11169-11186 (1996).

50. Kresse, G. & Furthmller, J. Efficiency of ab-initio total energy calculations for metals and semiconductors using a plane-wave basis set, *Computational Materials Science* **6**, 15-50 (1996).

51. Blochl, P. E. Projector augmented-wave method, *Physical Review B* **50**, 17953-17979 (1994).

52. Perdew, J. P., Burke, K. & Ernzerhof, M. Generalized gradient approximation made simple, *Physical Review Letters* **77**, 3865-3868 (1996).

53. Mostofi, A. A., *et al.* An updated version of wannier90: A tool for obtaining maximally-localised wannier functions, *Computer Physics Communications* **185**, 2309-2310 (2014).

54. Marzari, N., Mostofi, A. A., Yates, J. R., Souza, I. & Vanderbilt, D. Maximally localized wannier functions: Theory and applications, *Review of Modern Physics*, **84**, 1419-1475 (2012).

**Acknowledgements**

M.K., J.D., C.J., A.B., and E.R. performed the ARPES experiment and analyzed the resulting data. S.F. performed the theoretical calculations with help from H.C.P and E.K. L.Y. synthesized and characterized the single crystals. J.G.C. and R.C. supervised the project. M.K., S.F., and R.C. wrote the manuscript with input from all coauthors.

**Author contributions**

This work was supported by the STC Center for Integrated Quantum Materials, NSF Grant No. DMR-1231319. R.C. acknowledges support from the Alfred P. Sloan Foundation. This research was funded, in part, by the Gordon and Betty Moore Foundation EPiQS Initiative, Grant No. GBMF3848 to J.G.C. and ARO Grant No. W911NF-16-1-0034. M.K. acknowledges support from the Samsung Scholarship from the Samsung Foundation of Culture. L.Y. acknowledges support from the Tsinghua Education Foundation.

**Competing financial interests**

The authors declare no competing financial interests.

**Data availability**

The data that support the plots within this paper and other findings of this study are available from the corresponding author upon reasonable request.

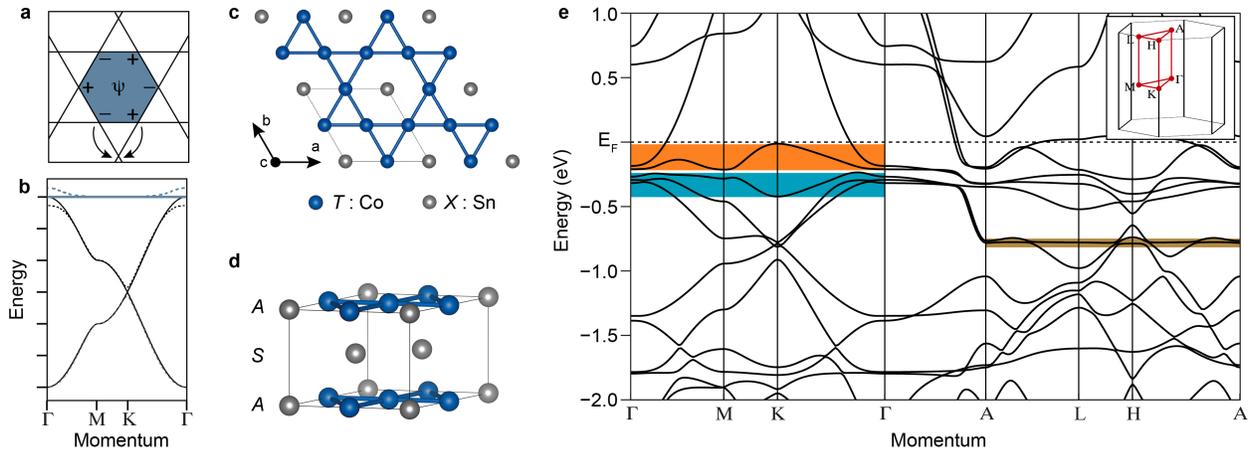

**Figure 1 | Electronic confinement and flat band in ideal kagome lattice and kagome metal CoSn. a,** Confinement of electron in kagome lattice with nearest neighbor hopping. Plus and minus signs indicate the phase of flat band eigenstate at neighboring sublattices. Any hoppings to outside of hexagon (arrows) are cancelled by destructive quantum interferences, resulting in the perfect localization of electron in the blue-colored hexagon. **b,** Tight-binding band structure of kagome lattice featuring flat band (blue solid line) and two Dirac bands with linear crossing at K (black solid lines). Inclusion of spin-orbit coupling gaps both Dirac crossing and quadratic touching point between the flat band and the Dirac band (Dotted lines). **c,** In-plane structure of kagome layer in CoSn consists of kagome network of Co atoms and space-filling Sn atoms. **d,** Three-dimensional structure of CoSn with alternating stacking of the kagome layer $A$ and Sn layer $S$. **e,** Relativistic density functional theory band structure of CoSn. Orange, cyan, and brown-colored regions highlight the manifestation of the kagome flat band in CoSn. Inset shows the bulk Brillouin zone of CoSn. The DFT band structure is shifted up by 140 meV to fit the experimental Fermi level.

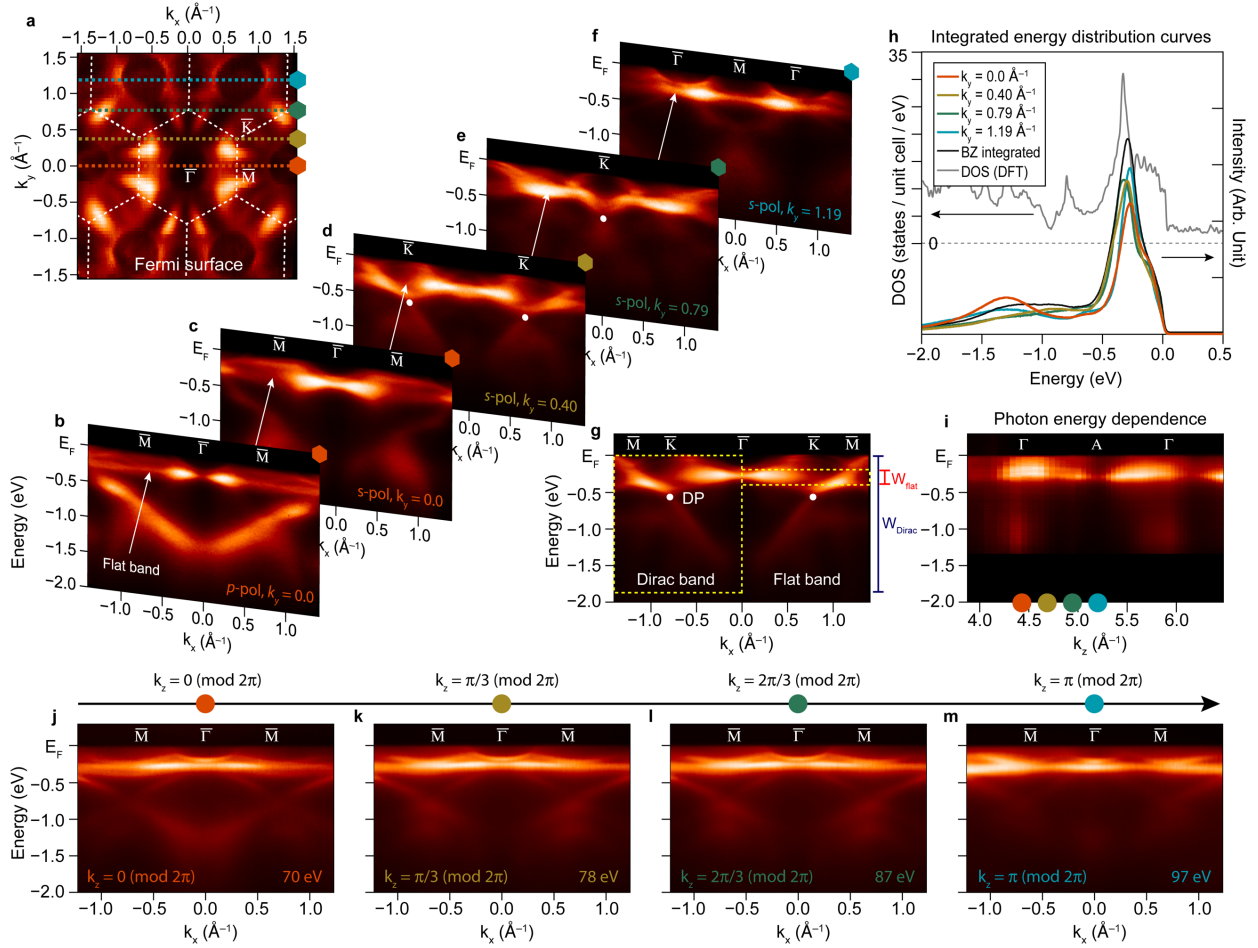

**Figure 2 | Direct visualization of flat band in CoSn**. **a,** Fermi surface of CoSn measured with 92 eV photons. Hexagonal surface Brillouin zones are marked with white-dashed lines. **b-f,** Energy-momentum dispersion of CoSn measured at $k_y = 0.0$ Å$^{-1}$ (b,c), 0.40 Å$^{-1}$ (d), 0.79 Å$^{-1}$ (e), and 1.19 Å$^{-1}$ (f). **g,** Energy-momentum dispersion of CoSn measured along $\bar{\Gamma} - \bar{K} - \bar{M}$ high-symmetry direction. The Dirac points at K are marked with white dots in c,d,g. Red and blue scales indicate the bandwidth of the Dirac and flat bands respectively. **h,** Momentum-integrated energy distribution curves from spectra in c-f (orange, brown, green, and cyan lines respectively), and from the entire first Brillouin zone (black line). Density of states curve from the density functional theory calculation of CoSn (grey line) is also overlaid. **i,** Out-of-plane dispersion of the flat band from photon-energy dependent ARPES experiment with the photon-energy tuned from 50 eV to 155 eV. **j-m,** Energy-momentum dispersion of the flat band along $\bar{\Gamma} - \bar{M}$ high-symmetry direction measured at representative $k_z = 0$, $\pi/3$, $2\pi/3$, and $\pi$ (mod $2\pi$) respectively.

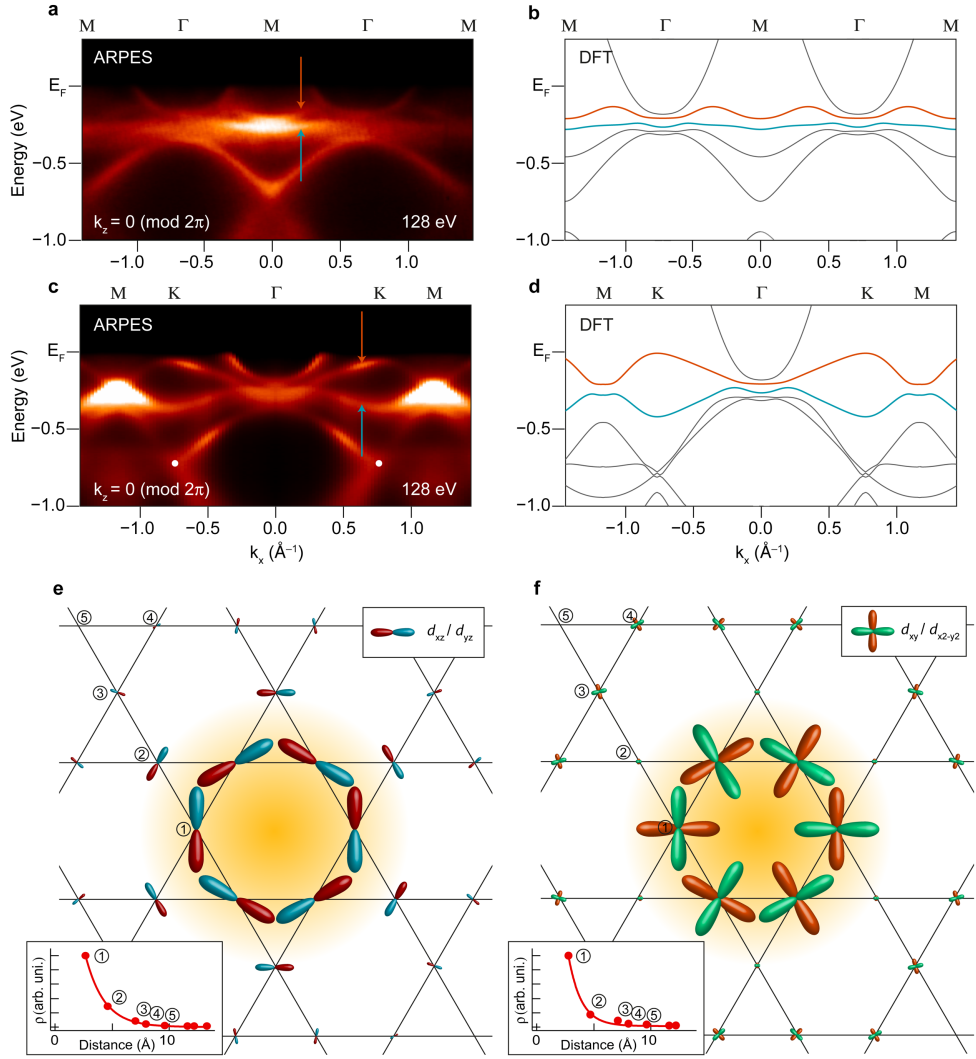

**Figure 3 | Orbital texture of flat bands and effective Wannier wave function. a,c,** High-resolution band structure of CoSn along Γ-M and Γ-K-M high-symmetry directions respectively. The data are measured with 128 eV photons which probes band structure at $k_z = 0$ plane. **b,d,** Corresponding DFT band structures. Cyan and orange lines respectively mark two flat bands arising from $d_{xz}/d_{yz}$ and $d_{xy}/d_{x2-y2}$ orbital degrees of freedom. Experimental dispersion of the flat bands (marked with orange and cyan arrows in a,c) are well-reproduced by the calculation. The Dirac point at K are marked with white dots in c. **e,f,** Orbital textures of the effective Wannier states constructed from the flat bands with $d_{xz}/d_{yz}$ and $d_{xy}/d_{x2-y2}$ orbitals respectively. Length scale of the orbitals at each site is proportional to the real part of orbital wave functions. Insets of e,f display the decay of total charge density (including the contribution from other orbitals) of the Wannier functions away from the central hexagon.

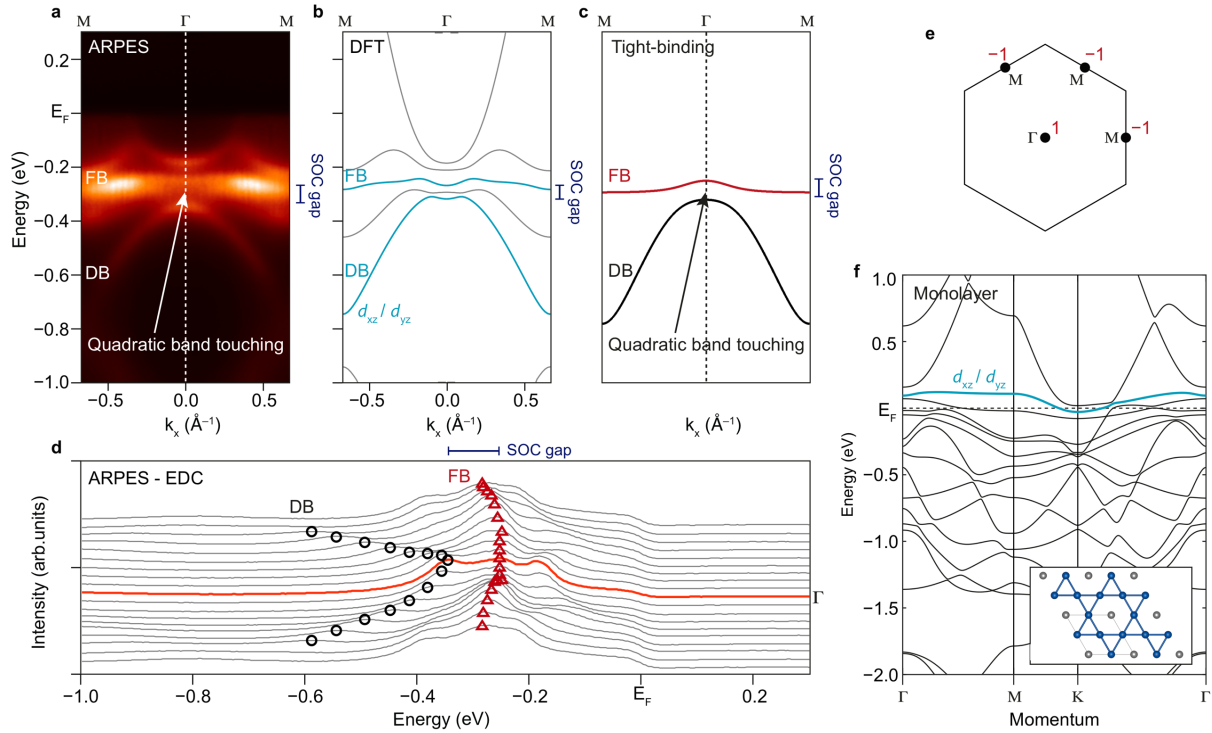

**Figure 4 | Spin-orbit-induced gap opening at the quadratic touching point between the Dirac band and the flat band. a,** High-resolution band structure of CoSn at the vicinity of the Γ point. In this image, we summed up the photoemission intensities from the first, second, and third Brillouin zone to minimize the effect of photoemission matrix element effect in intensity analysis. **b,c,** Corresponding DFT calculation of CoSn and *s*-wave tight-binding calculation of monolayer kagome lattice for direct comparison with ARPES spectrum in a. In b, we highlight the Dirac and flat bands with $d_{xz}/d_{yz}$ orbital characters with cyan color. **d,** Stack of energy distribution curves of the data in a. The dispersion of Dirac (flat) band with $d_{xz}/d_{yz}$ orbital character is tracked with black circles (red triangles). Energy distribution curve at the Γ point (marked with red line) displays the clear separation between the flat and Dirac band peaks, highlighting the gap opening at the quadratic band touching point. **e,** Parity eigenvalues at the time-reversal invariant momenta of $k_z$ = 0 plane of bulk CoSn. **f,** Electronic band structure for a single kagome layer limit of CoSn (inset). The $d_{xz}/d_{yz}$ flat band locates exactly at Fermi level and retains nonzero $Z_2$ index in the monolayer limit (parity eigenvalues are same with **e**).